# Certain aspects of prestack deconvolution

*Jagmeet Singh*
*CEWELL, ONGC,*
*Vadodara*

**Abstract**

In a previous paper, we had shown that because of varying angles of incidence there is a varying 'degree of convolution' down a trace and across a gather, necessitating deconvolution operators varying with time and offset. This idea is examined further in t-x as well as $\tau\text{-}p$ domain. We suggest better ways to deconvolve data in $\tau\text{-}p$ domain, taking into account varying degree of convolution in this domain. We derive formulae for periods of surface multiples in $\tau\text{-}p$ domain e.g. water column peg-legs and reverberations, which have a fixed period depending only on the value of $p$ —and suggest a way to check/revise the picked velocity using the formulae, provided the multiples are well separated from the primary. Periodicity of two way surface multiples is also studied.

**Introduction**

We had argued in an earlier paper (Singh, 2017) that since reflectivity series is sampled differently at different angles of incidence, a single (seismic) trace made up by different angles of incidence is not a simple convolution of the wavelet and reflectivity series. We further showed that even after slant stacking, the resulting $\tau\text{-}p$ trace in $\tau$-p domain $H(\omega, p)$ is not the product $W(\omega)R(\omega, p)$, but the product $W(\omega)R(\omega, k)$. We examine the consequences of this difference. We examine the idea of degree of convolution introduced earlier in great detail here, in both $t - x$ and $\tau - p$ domains

We study efficacy of $\tau\text{-}p$ deconvolution in removing surface multiples e.g. water column peg-legs. It is assumed that for a given p, these multiples are periodic, and that these can be addressed by $\tau\text{-}p$ deconvolution. We study this periodicity in different scenarios-- straight rays, ray bending at interfaces, and dipping beds. We take travel-times for these multiples of different orders, convert them to $\tau\text{-}p$ curves and study differences between these curves. We also study periodicity of two way surface multiples or double multiples (multiples that bounce back and forth between the surface and the primary reflector) in $\tau\text{-}p$ domain.

The paper is organised as follows. We first recapitulate briefly the results of our previous paper and then move on to examining how the $\tau\text{-}p$ transform affects convolution and degree of convolution in general, apart its effect on water column reverberations, peg-legs and two way surface multiples.

**Recap**

We recapitulate below the essential ideas presented in the paper by Singh (2017). The convolution sum or the trace value at a given time, looks like the following:

$$h(n) = w_0 R_n + w_1 R_{n-1} + w_2 R_{n-2} + \cdots \qquad (1)$$

for a given wavelet w(t) and reflectivity series R(t). The basic assumptions here are horizontal reflectors and normal incidence i.e. reflectors are sampled vertically. In reality, however, we have dipping reflectors and rays that are not incident normally, severely affecting the convolution sum (1). We consider the case of horizontal reflectors and slant rays appropriate for finite offset as in Figure 1. Two sets of rays at different depths are shown—we assume that the angle of incidence for rays within a set is the same, so that time difference between each pair of neighbouring rays within a set is assumed constant.

A unit of time in equation (1) is actually $\delta t_0$, zero offset time difference between neighbouring reflectors. Time difference at an offset can be found by differentiating NMO equation:

$$t^2 = t_0^2 + \frac{x^2}{v^2} \qquad (2)$$

We get

$$\delta t = \frac{t_0 \delta t_0}{\sqrt{t_0^2 + x^2/v^2}} \qquad (3)$$

or

$$\delta t = \delta t_0 cos\theta , \qquad (4)$$

where $\theta$ is the angle of incidence. We see that this difference shrinks with increasing $\theta$, making it necessary to introduce the idea of 'degree of convolution' [1]. Shallow reflectors imaged at high angles are more convolved (i.e. have more terms in the convolution sum) than deeper reflectors imaged at lower angles of incidence. The degree of convolution also increases with offset. So a deconvolution operator must change with time as well as offset. The idea of degree of convolution is explained well in Figure 2, where we see wavelets from three close reflectors well separated at lower offsets, but closer together at higher offsets.

Figure 3 shows two stacks with the section on the left showing the result of stacking gathers with four component surface consistent deconvolution, where the offset term is taken as the fourth term in the order of calculation and coarse offset classes are defined. The section on the right shows the result with the i) offset term taken as the first term in the order of calculation, ii) finer offset classes and iii) a second pass of just the offset term deconvolution(with the other terms suppressed). Note that to get a perceptible change in the output, all three steps had to be applied—though this may be data dependent.

We see a slight improvement in the second section showing that the offset term does take care of the offset dependence of deconvolution operator to some extent. Note that Cary and Lorentz (1993) have argued in favour of calculating the offset term, but not using it in filter application as ground roll gets collected in this term. We have been very careful in attenuating ground roll effectively using adaptive ground roll attenuation methods, so that we need not worry about this limitation. In order to take care of the time dependence of degree of convolution(and thereby the deconvolution operator), we carried out two window deconvolution with the same prediction distance(20 ms) and operator length(260 ms) for the two windows. Figure 4 compares the section on the right hand side of Figure 3(left hand side of Figure 4) with two window deconvolution (right hand side). We see a better stand-out of events in the lower parts of the section.

**τ- p domain**

Formula (4) above suggests that it should be advantageous to go to $\tau\text{-}p$ domain, where degree of convolution should be constant for a given $p$ or $\theta$, if velocity is constant. Our discussion above shows that reflectivity series should be considered not only a function of time t, but also that of offset x. We showed in [1] that when the substitution

$$t = \tau + px \qquad (5)$$

is made in the convolution integral:

$$h(t,x) = \int W(t-t')R(t',x)dt', \qquad (6)$$

and summation made over x, we get the result:

$$H(\omega,p) = W(\omega)R(\omega,k) \qquad (7)$$

With argument p on L.H.S. and k on R.H.S. of the above equation, the natural question that arises is, 'is the $\tau, p$ trace really a convolution of the wavelet and reflectivity at the particular p as assumed, or not?'. We get $R(\omega, k)$ on the R.H.S. instead of $R(\omega, p)$ simply because sampling of the reflection series is constant for a given angle (which in turn depends on $k_x = p\omega$ and $k_z = q\omega$) and not for a given $p$. In our case, however, since we assume constant velocity, constant $p = \frac{sin\theta}{v}$ implies constant $\theta$ as well, so there is really no difference. That is, $for\ all\ \omega$, angle related to $k_x/k_z$ is constant for a given $p$. Degree of convolution dependent on angle should also be the same for all $\tau$ for a given $p$ trace for a constant velocity medium. Note however that each amplitude from t-x domain, entering into $\tau\text{-}p$ transform, is already convoluted (differently at different times and offsets) and this effect cannot be undone

For the more realistic case of vertically varying velocity, angle would change at every interface affecting degree of convolution down a $p$ trace. So we suggest using time varying deconvolution operators even in the case of $\tau, p$ deconvolution. Another route, we suggest is to wavelet transform a $\tau, p$ trace, so that $k_x = \omega p$ or angle is constant for a given $\omega$ implying a constant degree of convolution down a $p$ trace. So, the alternate route suggested is to deconvolve every wavelet-transformed $\omega - p$ trace separately, and reconstruct back the deconvolved $\tau, p$ gather.

With the question of deconvolution in $\tau\text{-}p$ domain settled, we now come to the next question: periodicity of surface multiples for a fixed $p$, justifying $\tau\text{-}p$ deconvolution as a good option for removal of multiples. We consider below two different type of surface multiples—water column multiples and double multiples.

**Multiple periodicity in $\tau$-$p$ domain.**

We consider an important type of multiple: primaries followed by their reverberations in the water column. A primary given by say $t = \sqrt{t_0^2 + \frac{x^2}{v^2}}$ and its reverberations $t = \sqrt{t_0^2 + \frac{x^2 t_0^2}{(vt_0+nv't_0')^{\wedge}2}} + n\sqrt{t_0'^2 + \frac{x^2 t_0'^2}{(vt_0+nv't_0')^{\wedge}2}}$, where n=1,2,3 etc. represents the order of the reverberation, $t_0'$ the two way vertical time in the water column and $v'$ the water velocity, are shown in Figure 5 in x-t domain (for $t_0 = 3s$, $t'_0 = 17.32\ ms$ and $v = 3\ km/s$. $v' = 1.5\ km/s$ ). These equations can be easily derived for the primary and multiples at a given $x$, using straight ray paths. Let us examine if these can be addressed by slant stacking followed by $\tau - p$ deconvolution. In order to transform these to $\tau - p$ domain, we need to draw tangents of different slopes $p$ to all the curves and find corresponding $\tau$ i.e. intercepts on y-axis. Curves obtained corresponding to Figure 5 in $\tau - p$ domain are shown in Figure 6. We see that these multiples are almost periodic in (x,t) as well as $\tau$-$p$ domain. To study this further, we increase the value of $t'_0$ to 1.5 s keeping other parameters the same, so that the time differences, if any, can be accentuated. We can see from Figures 7-10 that the reverberations become aperiodic at large offsets in (x,t) domain and also at higher values of $p$ in $\tau$-$p$ domain.

Now we come to the case of multiples that reverberate between the free surface and the primary reflector. We call these simply double multiples, though this may be a misnomer for $2^{nd}$ order double multiple (as it would really be a triple multiple). We consider the primary $t = \sqrt{9 + \frac{x^2}{9}}$ at $t_0 = 3s$ for $v = 3\frac{km}{s}$, with two of its surface multiples given as $t = \sqrt{36 + \frac{x^2}{9}}$ and $t = \sqrt{81 + \frac{x^2}{9}}$. Figure 11 shows $\tau - p$ curves corresponding to the primary and its first and second order multiples. We have included the grid so that differences between curves may be estimated. We have also plotted the difference curves (i.e. difference between $1^{st}$ order multiple and primary, between $2^{nd}$ order and $1^{st}$ order multiples of Figure 11) in Figure 12, and we see that the double multiples are exactly periodic. So, can we expect double multiples to be well attenuated by $\tau - p$ deconvolution? In principle yes, but two way surface multiples are generally long period multiples and according to Xiao et. al. (2003), deconvolution may not be the best choice to attenuate these 'because there may not be enough multiples, in the record length to satisfy the periodic requirements. Another problem is that long–period multiples require long operators. Since primaries can be periodic over long time windows, long operators have the potential to suppress primaries as well as multiples'(quoted from Xiao et. al.(2003)).

We now try to understand intuitively why water column multiples have turned out to be non-periodic in $\tau - p$ domain while two way surface multiples turn out periodic. A hyperbola for

offset x, angle of incidence $\theta$, velocity v and zero offset time $t_0$, we may write two way time t as

$$t = t_0 \sec\theta \ \& \ x = vt_0 \tan\theta, \qquad (8)$$

so that

$$p = \frac{dt}{dx} = \frac{dt}{d\theta}\frac{d\theta}{dx} = \frac{\sin\theta}{v}, \qquad (9)$$

as expected. First order water column reverberation for offset x, primary reflector at depth $vt_0$ and water depth = $v't'_0$ is shown in Figure 13 (Figure 14 for the OBC case). We see that now, we may write two way time t as

$$t = t_0 \sec\theta + t'_0 \sec\theta \qquad (10)$$

and

$$x = vt_0 \tan\theta + v't'_0 \tan\theta. \qquad (11)$$

From equations (10) and (11), we get

$$\frac{dt}{dx} = \frac{dt}{d\theta'}\frac{d\theta'}{dx} = \frac{(t_0+t'_0)\sin\theta}{vt_0+v't'_0} \qquad (12)$$

Equation (12) can be equated to equation (9) to find out where the values of p match, and the corresponding $\tau$'s found by drawing tangents and finding the intercepts. Though cumbersome, the calculation is doable and we have already seen from Figures (6-10) that the resulting time differences are not periodic (though almost periodic).

If we take $v = v'$ in Equation (12), p reduces to the familiar expression in Equation (9). It is now much easier to calculate the intercepts and we find $\tau = t_0\cos\theta, (t_0 + t'_0)\cos\theta, (t_0 + 2t'_0)\cos\theta, ..$ etc. for the primary, 1$^{st}$ order multiple, 2$^{nd}$ order multiple etc. implying a fixed period of $t'_0\cos\theta$ between 2 consecutive order multiples-- as we have seen for the double multiples (Figure (11-12), where $t_0 = t'_0$ as well). Further note that the value of p at a given x for the primary is found at 2x, 3x etc. on the 1$^{st}$ order multiple, 2$^{nd}$ order multiple, etc. respectively.

So are the water column peg-legs really aperiodic? So far, we have been assuming straight rays and neglected ray bending or refraction effects at interfaces. It would be prudent to calculate $\tau - p$ curves, allowing for ray bending at the medium-water column interface. In this case, we have

$$\frac{\sin\theta}{v} = \frac{\sin\theta'}{v'}, \qquad (13)$$

where $v$ is the velocity of the medium and $v'$ the velocity within water. Using $t_0$ and $t'_0$ as defined above, we can show that equations corresponding to (10) and (11) above take the form

$$t = t_0\sec\theta + \frac{t'_0 v}{\sqrt{v^2-v'^2\sin^2\theta}} \qquad (14)$$

and

$$x = vt_0 \tan\theta + v't'_0 \frac{v'\sin\theta}{\sqrt{v^2 - v'^2 \sin^2\theta}} \quad (15)$$

Using equation (14) and (15), it can be shown that

$$p = \frac{dt}{dx} = \frac{\sin\theta}{v}, \quad (16)$$

which must be true as the value of p is preserved on refraction. Furthermore using

$$\tau = t - px, \quad (17)$$

with values inserted from Equations (14-16), we get the $\tau - p$ curves of Figure (15), and obtain perfect periodicity at all values of p, as is evident from the difference curves of Figure (16). So whatever aperiodic behaviour we saw in Figures (6-10) was because of ignoring refraction or Snell's law, essential for preserving the value of p. Furthermore, it is easy to work out the period $\tau_2 - \tau_1$ (using Equations 14-17), between the primary and first multiple or between two consecutive water column multiples and is as follows:-

$$\tau_2 - \tau_1 = \frac{t'_0 v}{v'\sqrt{v^2 - v'^2 \sin^2\theta}} - \frac{v'^2 t'_0 \sin\theta}{v\sqrt{v^2 - v'^2 \sin^2\theta}} \quad (18)$$

Since parameters $t'_0$ related to water depth, and acoustic velocity in water $v'$ in above equation are known, this gives us a tool to check our picked velocity $v$ from the period seen on $\tau - p$ curves, provided the water column multiple is at a discernible interval from the primary.

**Dipping beds**

Whatever we have discussed above is valid for horizontal reflectors. The moment we add dip, the situation becomes much more complicated. As the simplest case, we consider a single dipping bed with dip $\alpha$ in Figure (17). Travel time equation for the dipping interface is given as:-

$$t^2 = t_0^2 + \frac{x^2 \cos^2\alpha}{v^2} \quad (19)$$

Angle of incidence is taken as $\theta$. Differentiating the above equation and noting that zero offset time $t_0$ is itself a variable i.e. changes with $x$ or $\theta$, we obtain

$$t\,dt = t_0 dt_0 + \frac{x \cos^2\alpha \, dx}{v^2} \quad (20)$$

Using the simple geometrical relation $dt_0 = dx \sin\alpha$, we get

$$\frac{dt}{dx} = \frac{t_0}{t}\sin\alpha + \frac{x\cos^2\alpha}{v^2} \quad (21)$$

Further using sine rule for triangles, we can see that

$$\frac{x}{\sin\theta} = \frac{vt}{\cos\alpha} \quad (22)$$

Substituting (22) in (21), we get

$$\frac{dt}{dx} = \frac{sin\alpha cos\theta}{v} + \frac{cos\alpha sin\theta}{v} = \frac{sin(\alpha + \theta)}{v} \qquad (23)$$

We see that angle $(\alpha + \theta)$ is the angle that RO makes with the vertical in Figure 17. Now if we introduce a water column of fixed depth above the surface in Figure 17, the value of $p = \frac{dt}{dx}$ will not change in the water column, and it can be easily shown along the lines of Equation (18), that the period of water column peg-legs is fixed in $\tau - p$ domain. But as more beds with variable dips or even flat beds are added (above or below this bed), the situation becomes more complicated and $p = \frac{dt}{dx}$ would not be of the simple form given in Equation (23)—thereby the value of p in the water column will not be preserved implying that the period of peg-legs will not be fixed, making deconvolution ineffective in addressing these.

**Conclusions**

We introduced the idea of degree of convolution and saw how it necessitated deconvolution operators changing with time and offset. We suggested using time varying deconvolution operators even in the case of $\tau, p$ deconvolution, or alternatively to wavelet transform a $\tau, p$ trace (so that $k_x = \omega p$ or angle would be constant for a given $\omega$ implying a constant degree of convolution down a $p$ trace), deconvolve every wavelet-transformed $\omega - p$ trace separately, and reconstruct back the deconvolved $\tau, p$ gather. That is, deconvolution with either time varying operators in $\tau, p$ domain, or deconvolution in wavelet-transformed $\tau, p$ domain or $\tau - p - \omega$ domain should be the way to go.

From analysis of curves in $\tau - p$ domain, we conclude that the assumption that water column peg-legs are periodic in the $\tau - p$ domain i.e. they have a fixed period for a fixed p, is correct only if refraction or ray bending effects at interfaces are considered. If these effects are ignored, we are led to the wrong conclusion that that water column multiples are aperiodic in $\tau - p$ domain. Two way surface or double multiples are exactly periodic for any value of p, suggesting good elimination by $\tau - p$ deconvolution (in principle, at least). Water column peg-leg suppression using $\tau - p$ deconvolution is a viable option as long as geology is not complex i.e. beds are more or less flat.

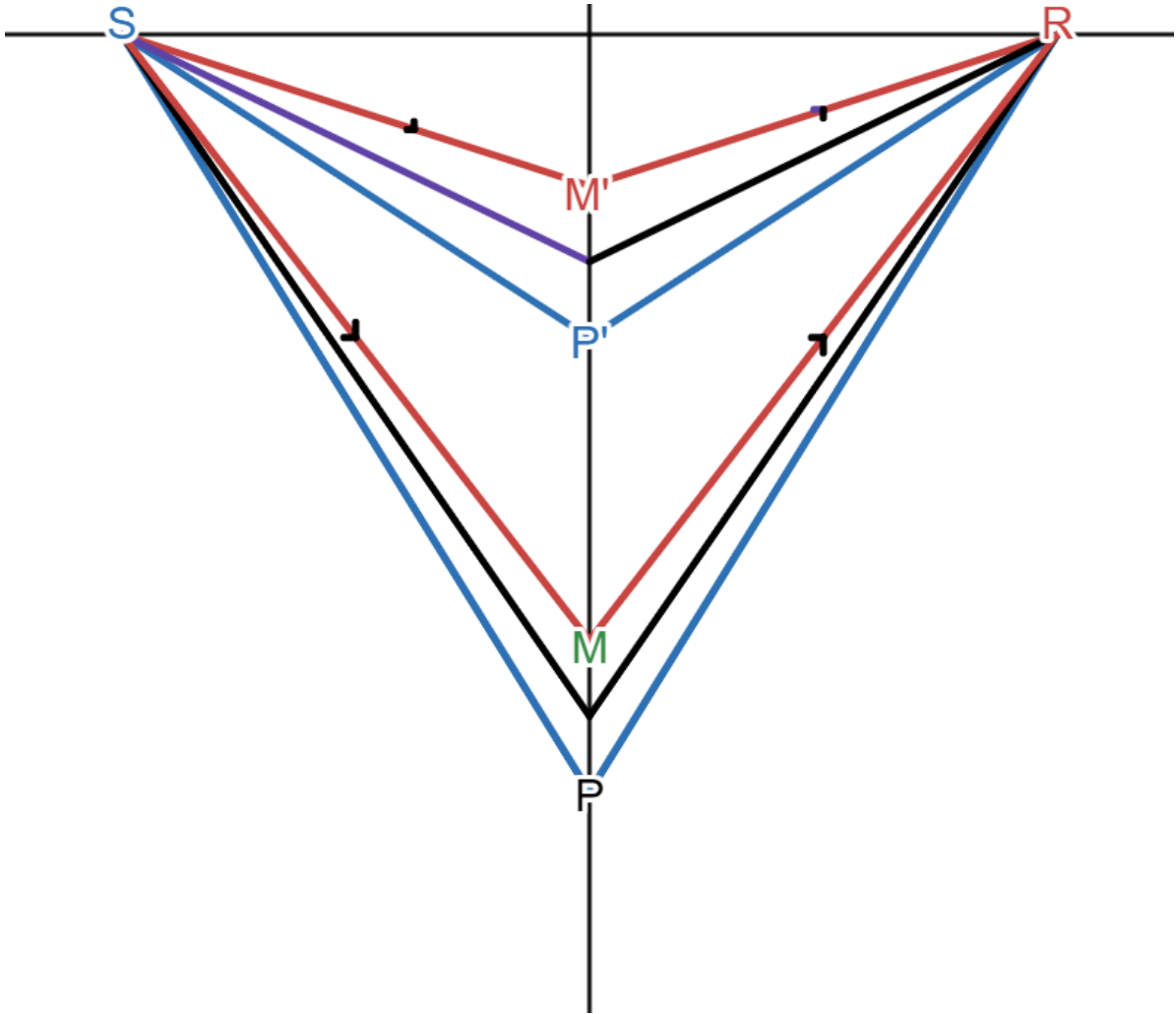

Figure 1 : Two sets of rays (incident and reflected) at different depths

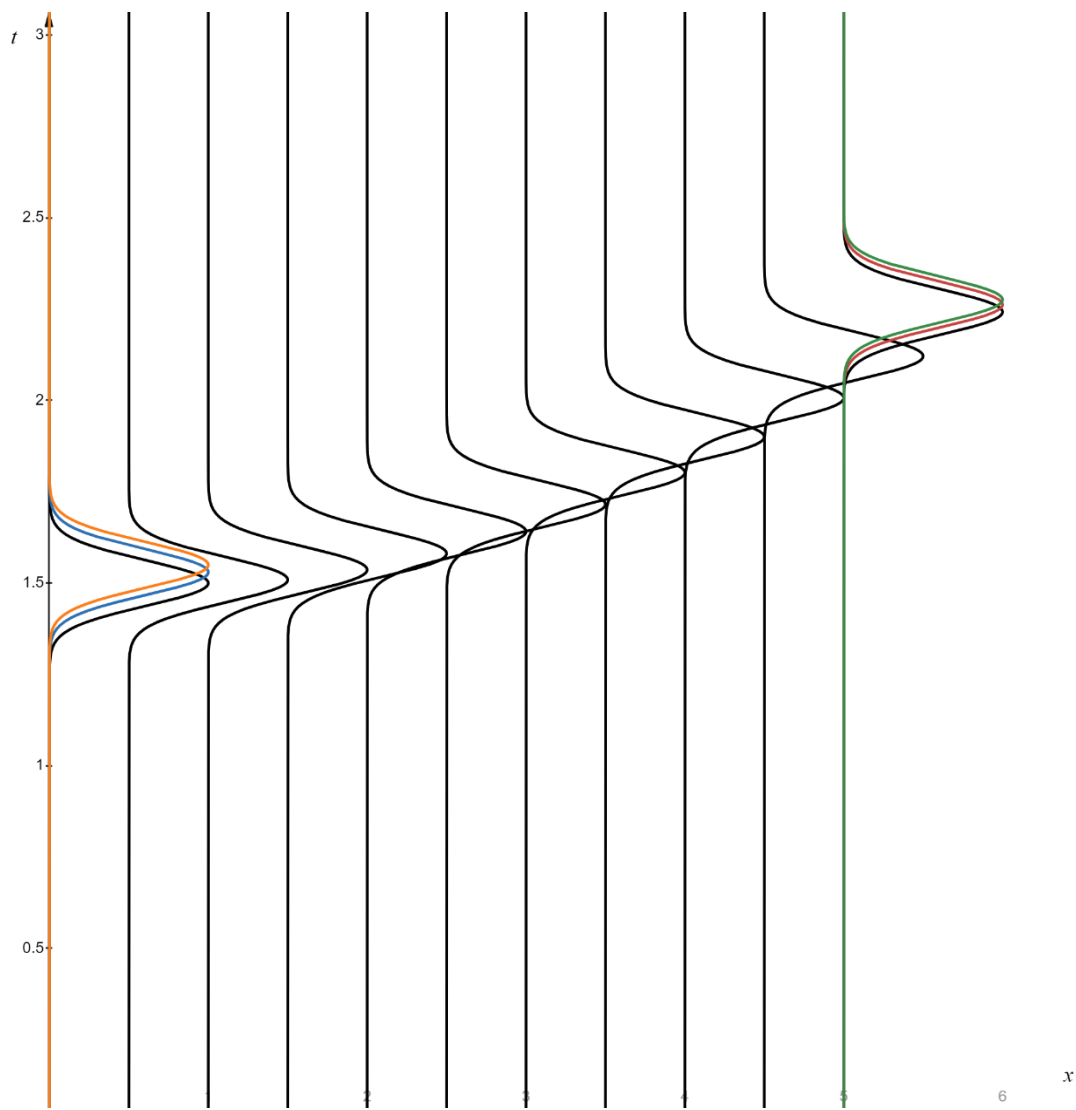

Figure 2: The idea of degree of convolution is explained well here—wavelets from close by reflectors are well separated at lower offsets, whereas they are closer together at higher offsets (implying a higher degree of convolution).

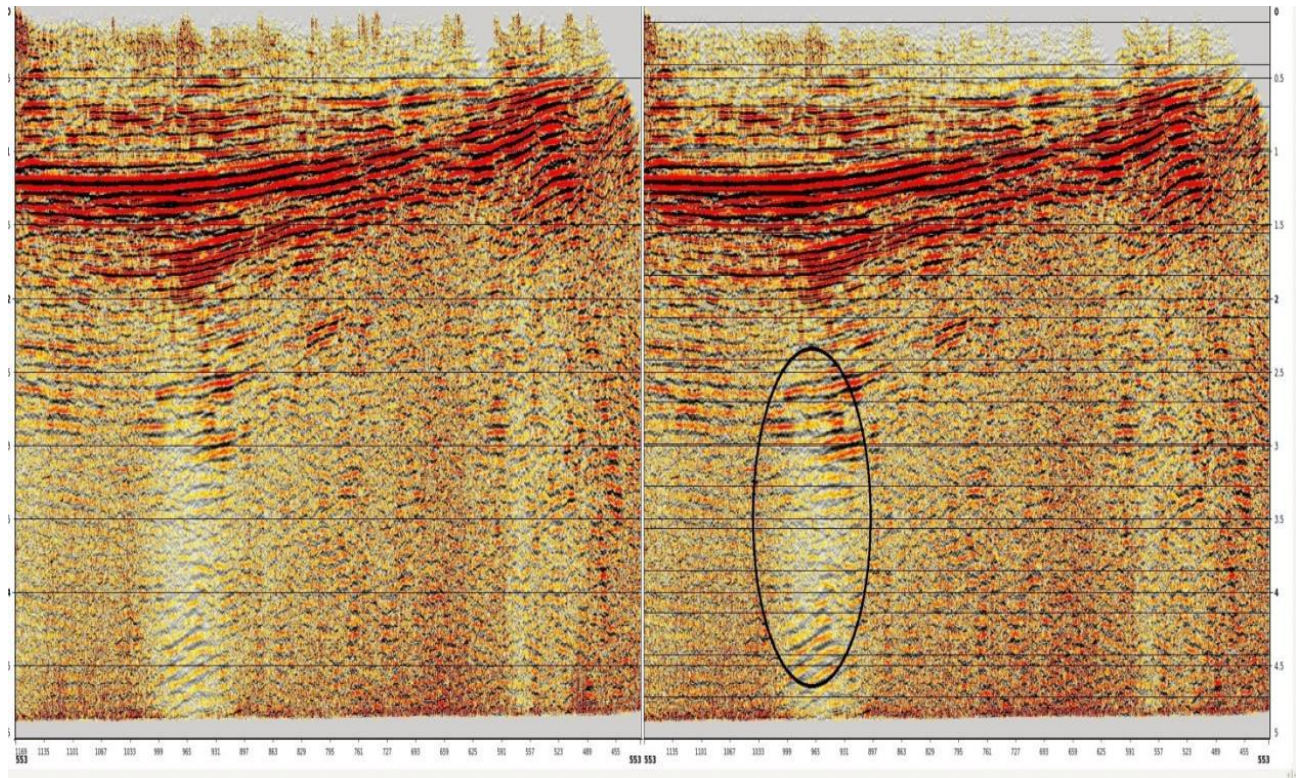

Figure 3: The section on the left shows the result of four component surface consistent deconvolution where the offset term stands fourth in the order of calculation (compared to section on the right which uses the offset term as the first term to be calculated).Coarser offset classes are used in the section on the left compared to the section on the right.

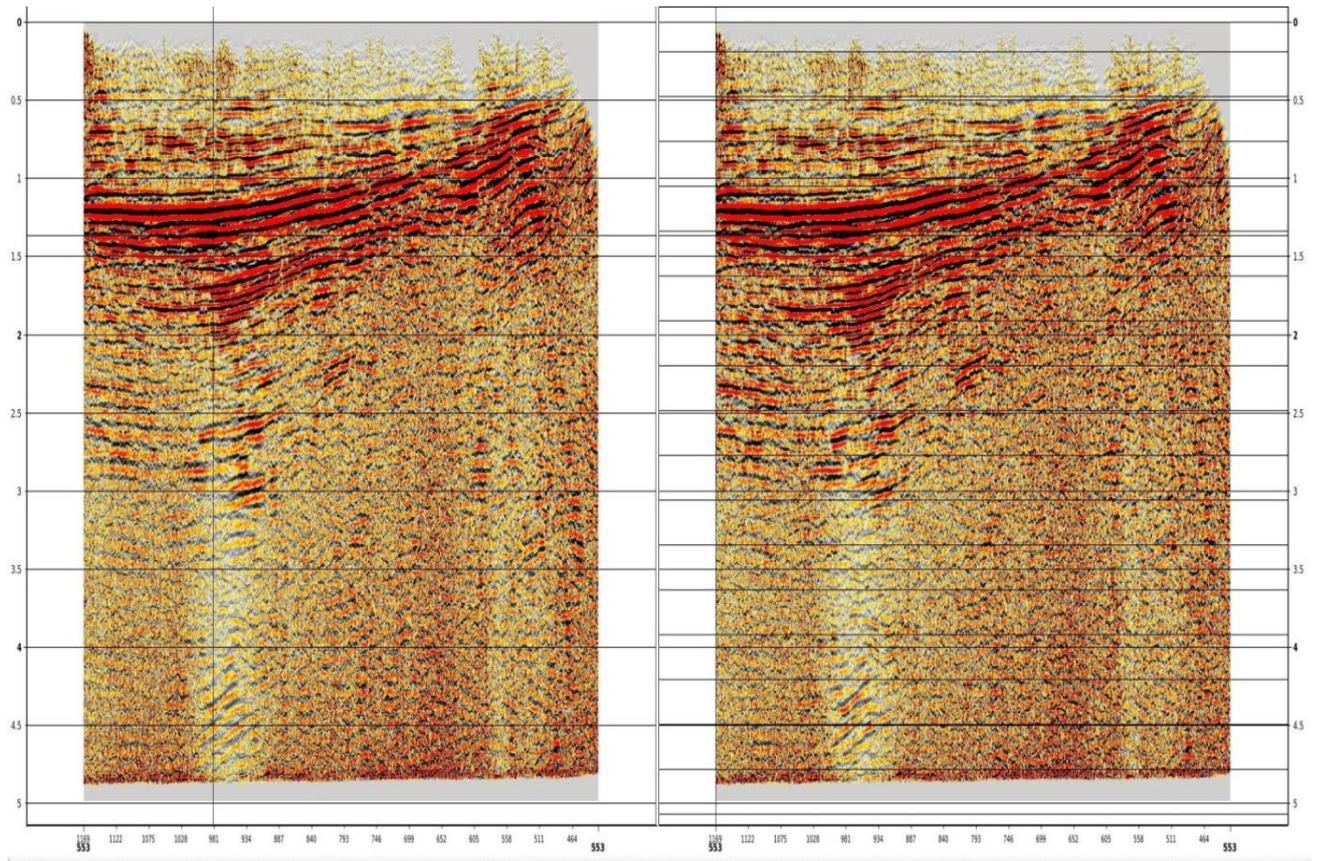

Figure 4: Further improvement seen with two window deconvolution with same P.D. (20 ms) and O.L. (260 ms)

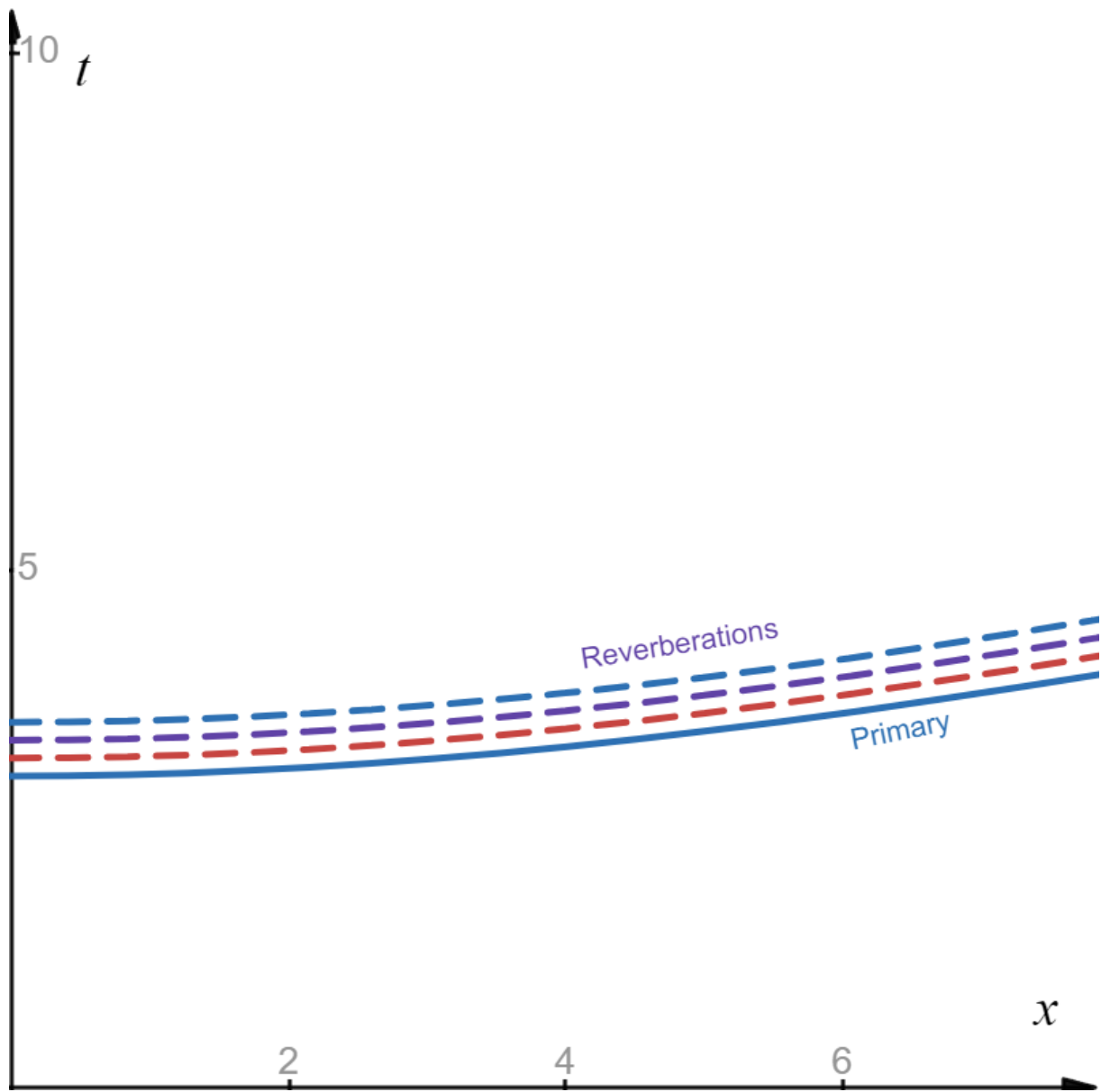

Figure 5: A primary and its peg-legs in the water column for $t_0 = 3s$, $t'_0 = 17.32\ ms$ and $v = 3\ km/s$. $v' = 1.5\ km/s$ ( water depth $= v't'_0 = 25.98m$)

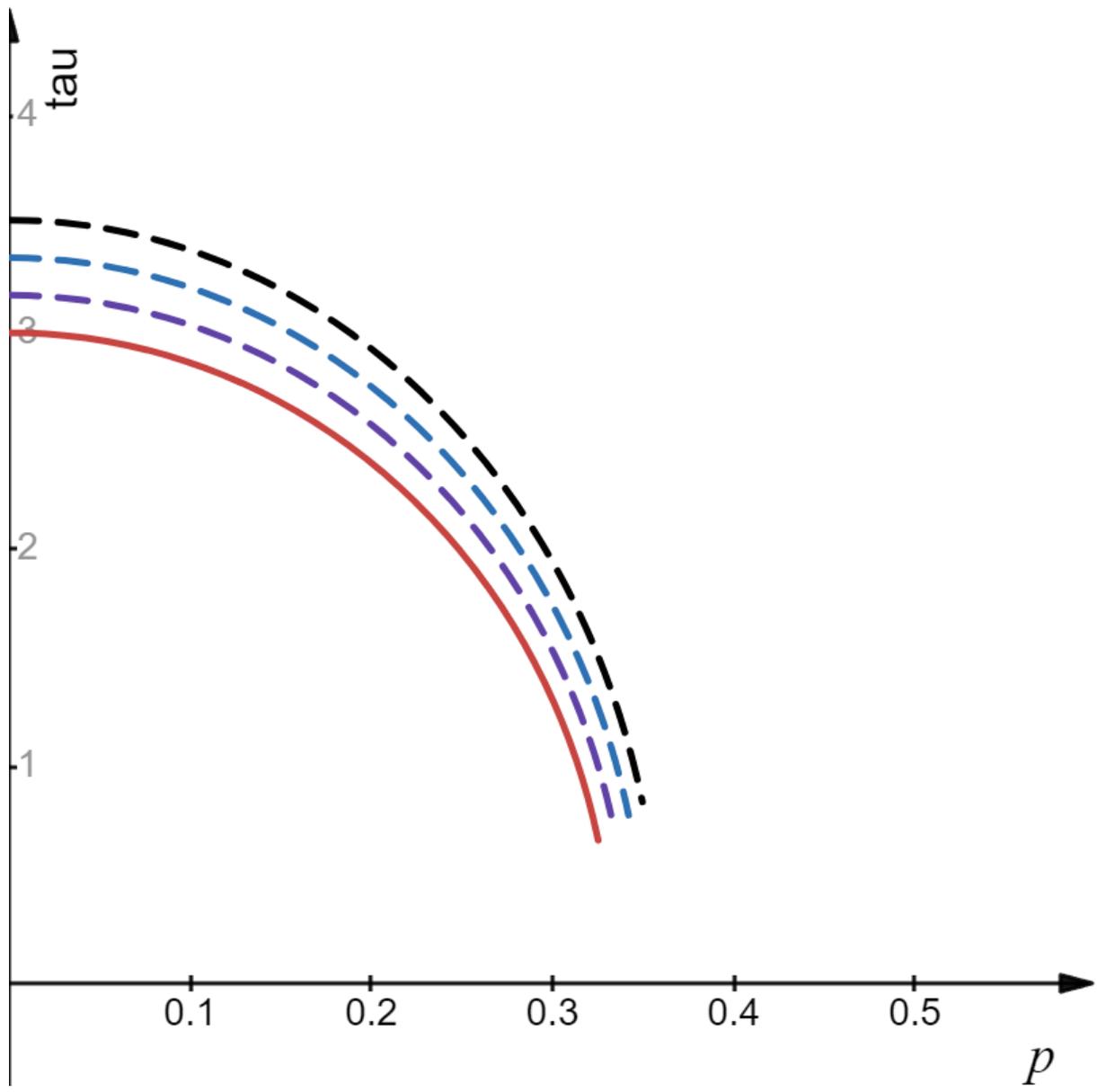

Figure 6: Primary and multiples of figure 5 in $\tau$-$p$ domain.

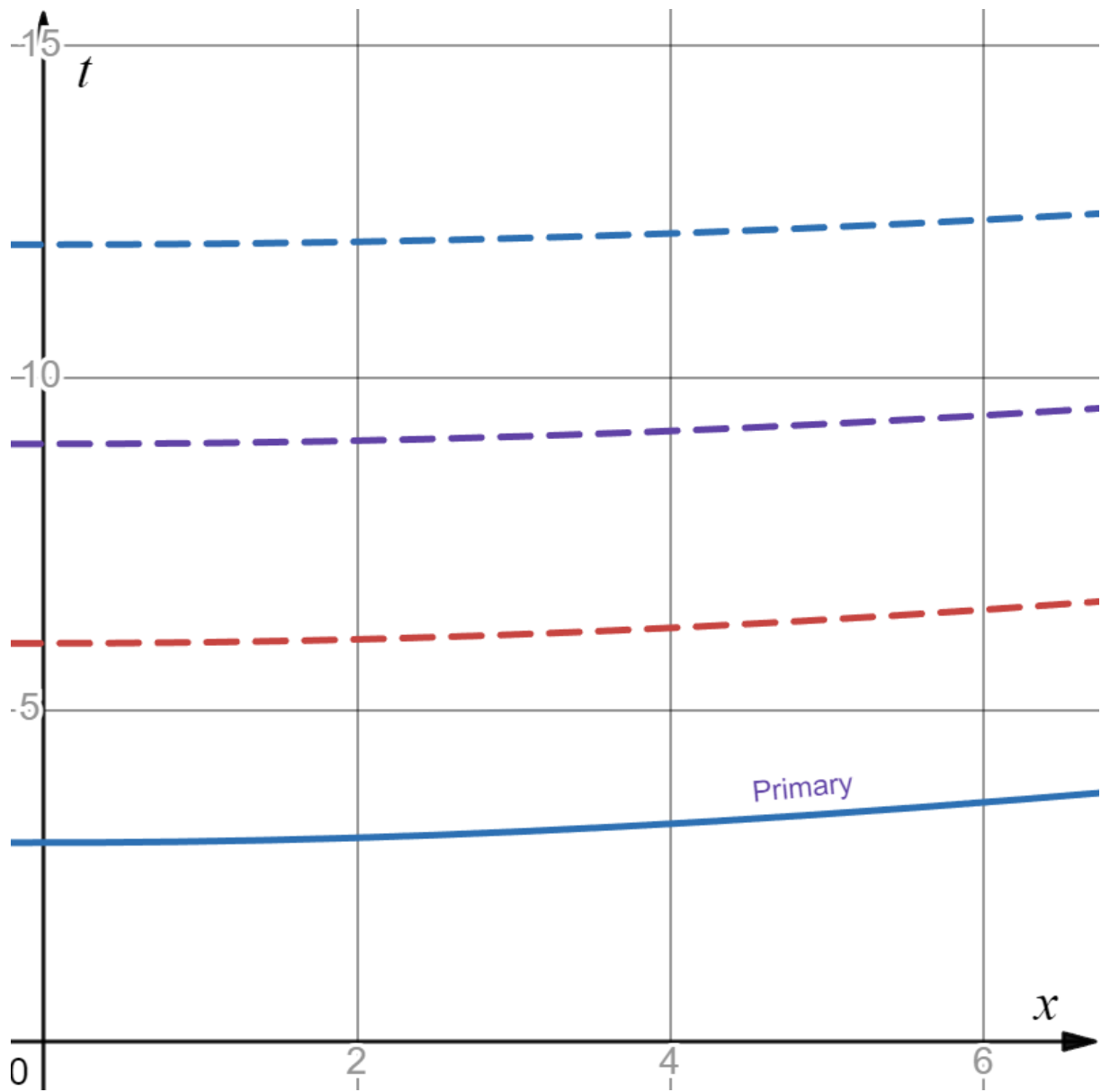

Figure 7: Primary and water column peg-legs for $t_0 = 3$s $t'_0 = 1.5$s and $v = 3$ km/s. $v' = 1.5$ km/s. Multiples seem almost periodic for lower values of offset x, becoming aperiodic at large offsets.

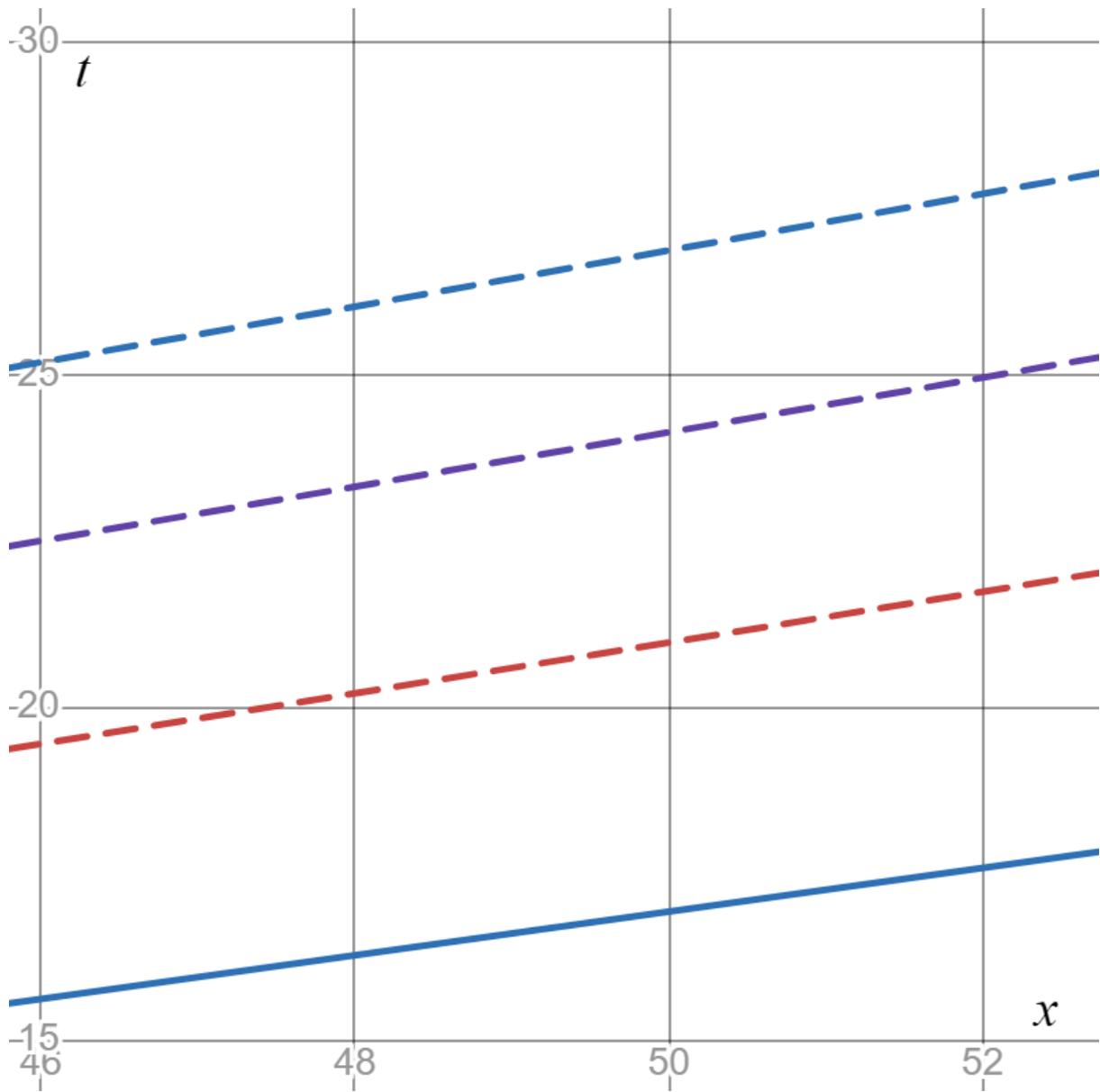
Figure 8: Primary and multiples of Figure 7 shown at large offsets.

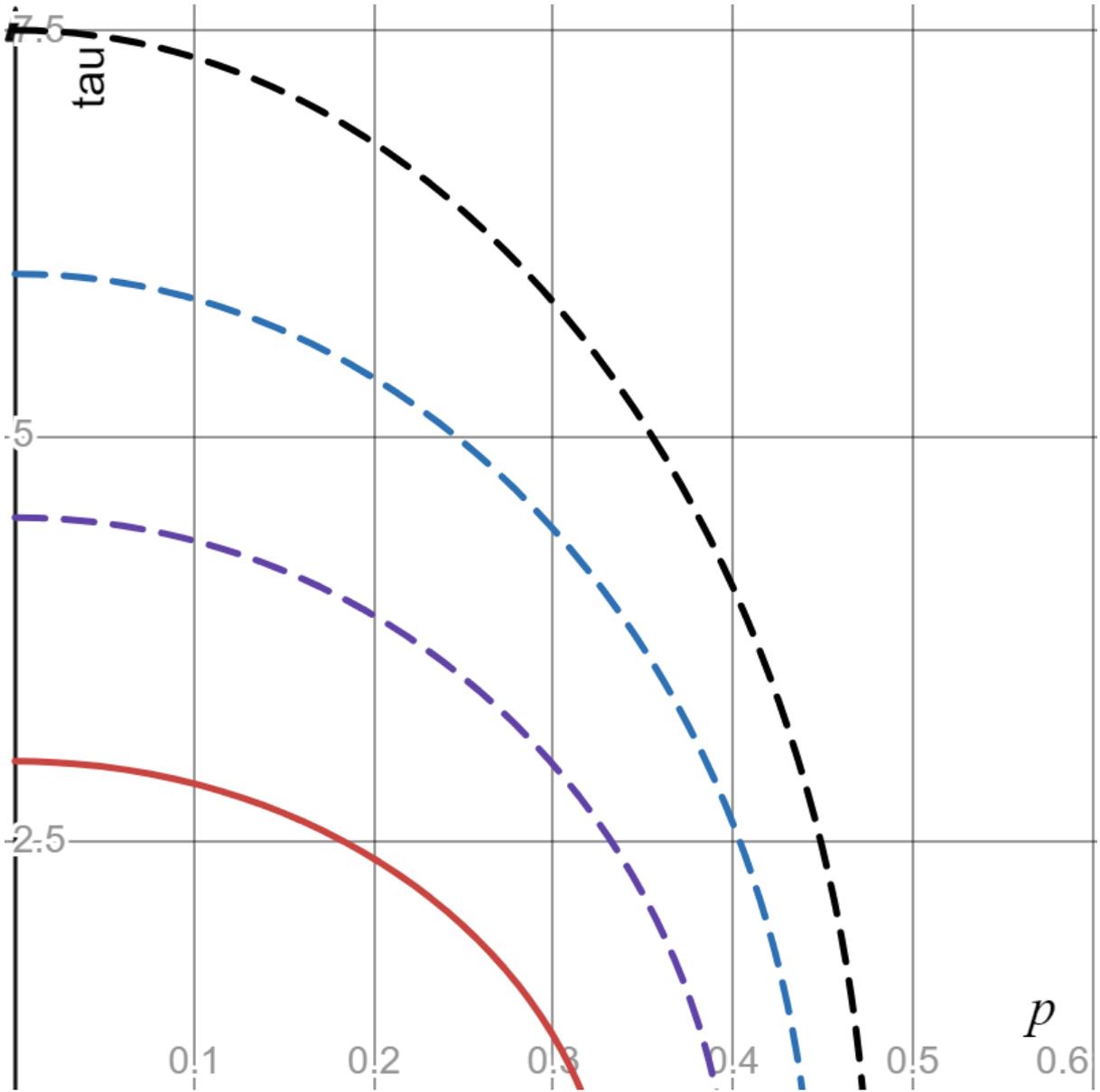

Figure 9: $\tau - p$ curves corresponding to figure (7).

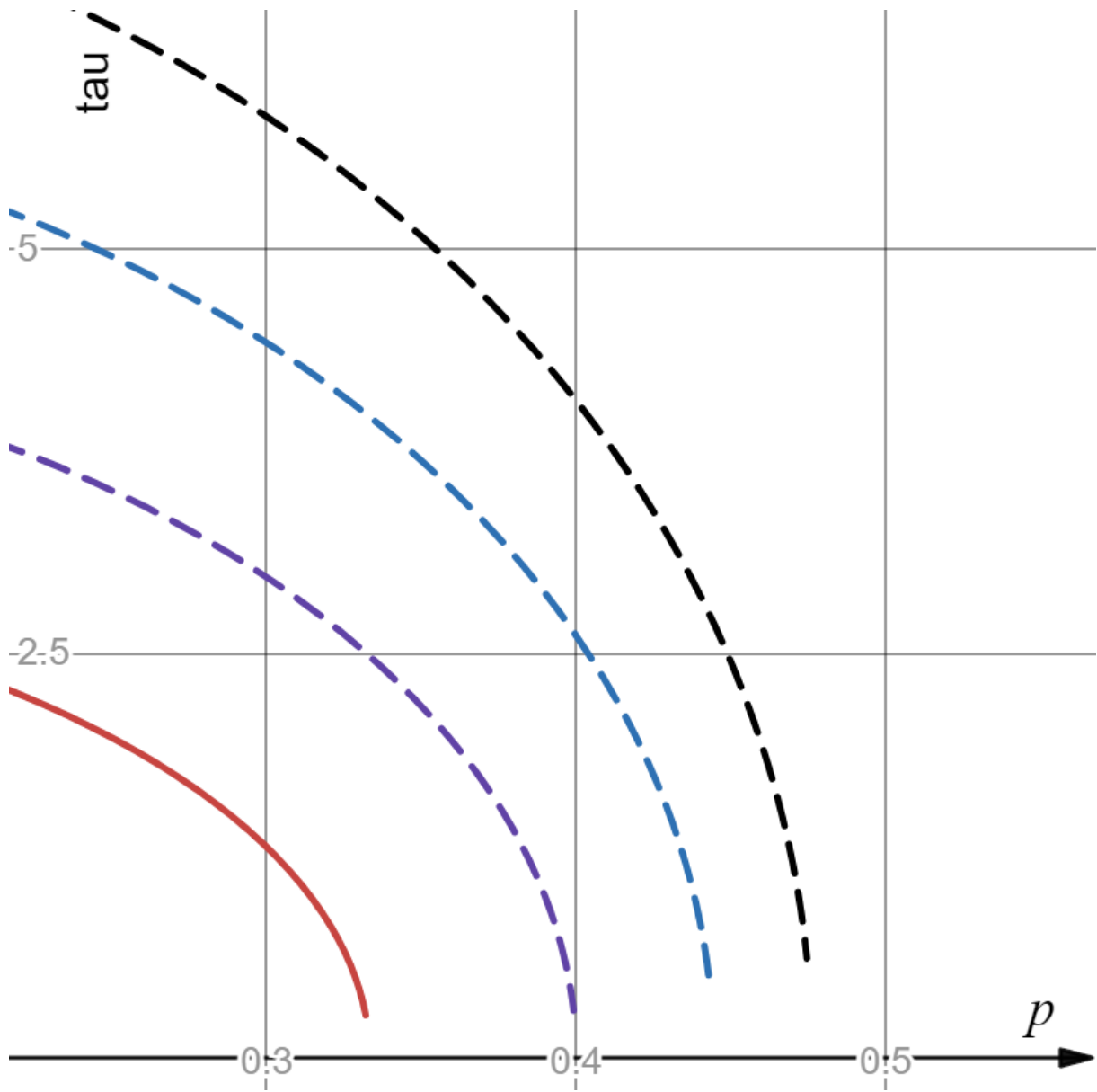

Figure 10: $\tau - p$ curves of Figure 9 at higher values of p. We see that reverberations are clearly aperiodic..

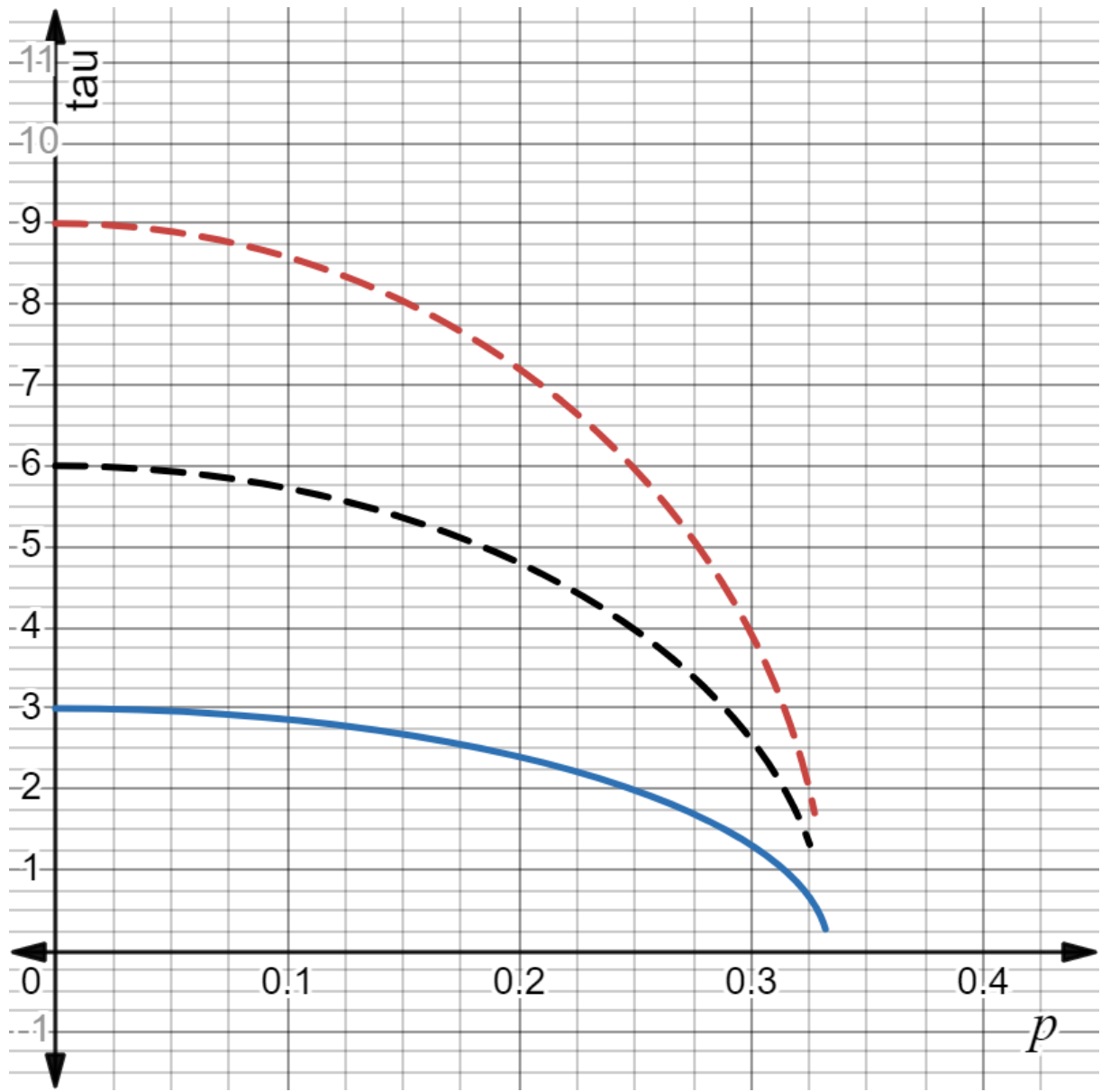

Figure 11: $\tau$- $p$ curves corresponding to primary (blue) and its first and second order double multiples.

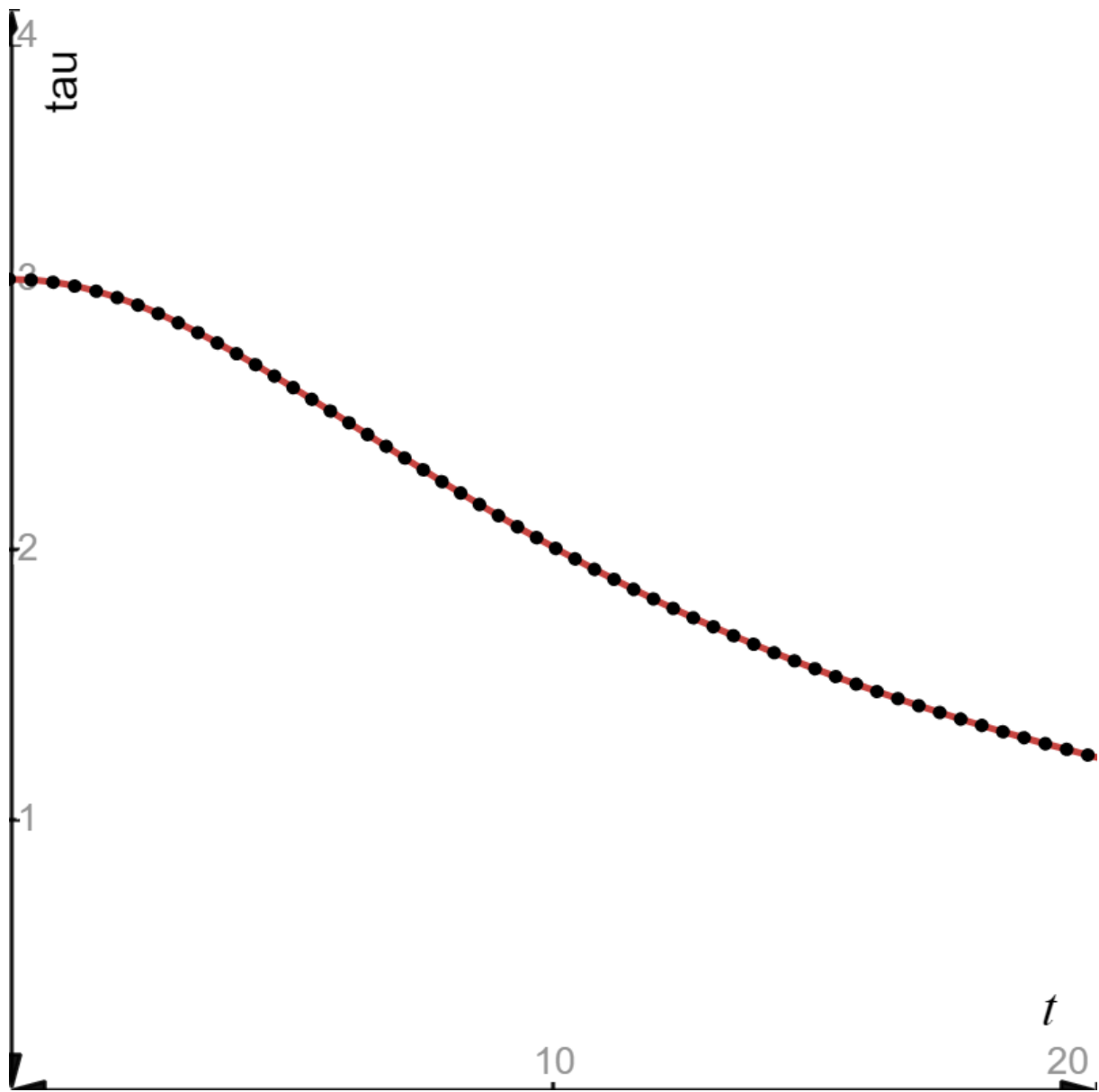

Figure 12: Difference between 1st order surface multiple and primary (black dotted curve) and difference between 2nd order and 1st order double multiples (red curve) of figure (11). 't' is an intermediate parameter used to define p in a parametric equation.

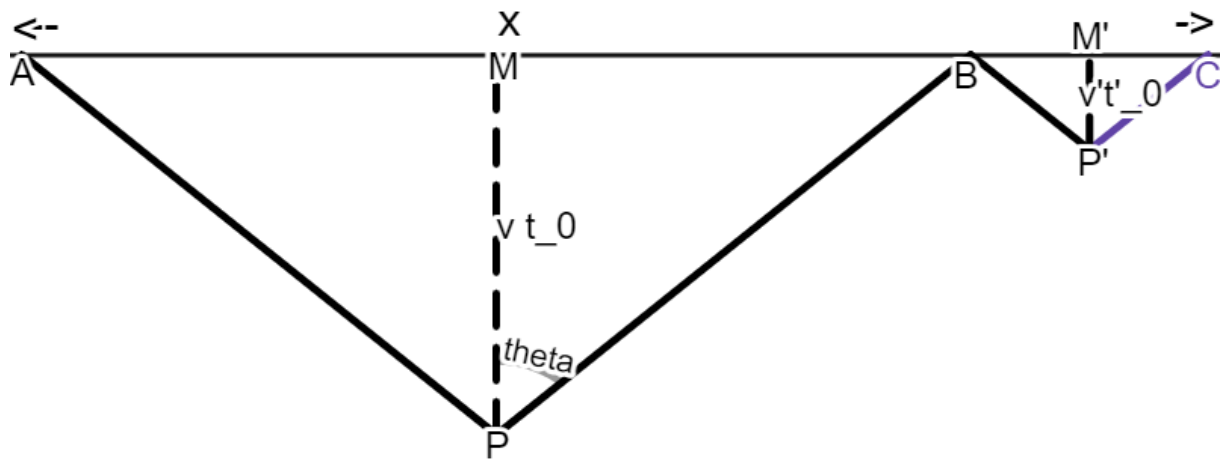

Figure 13: First order water column peg-leg for offset x = AC, primary reflector at depth $vt_0$ and water depth = $v't'_0$

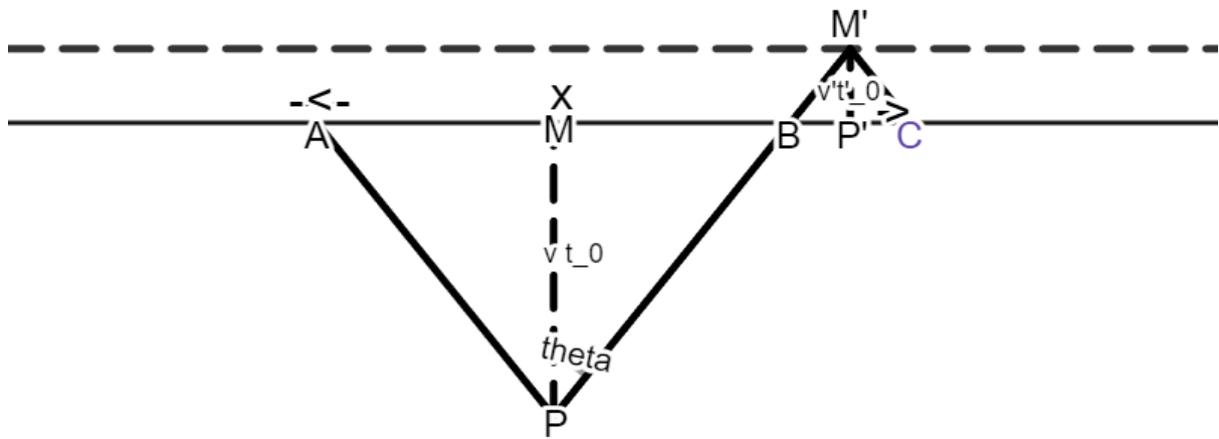

Figure 14: First order water column peg-leg as in figure 13, where we take water column above the receivers (as in OBC case).

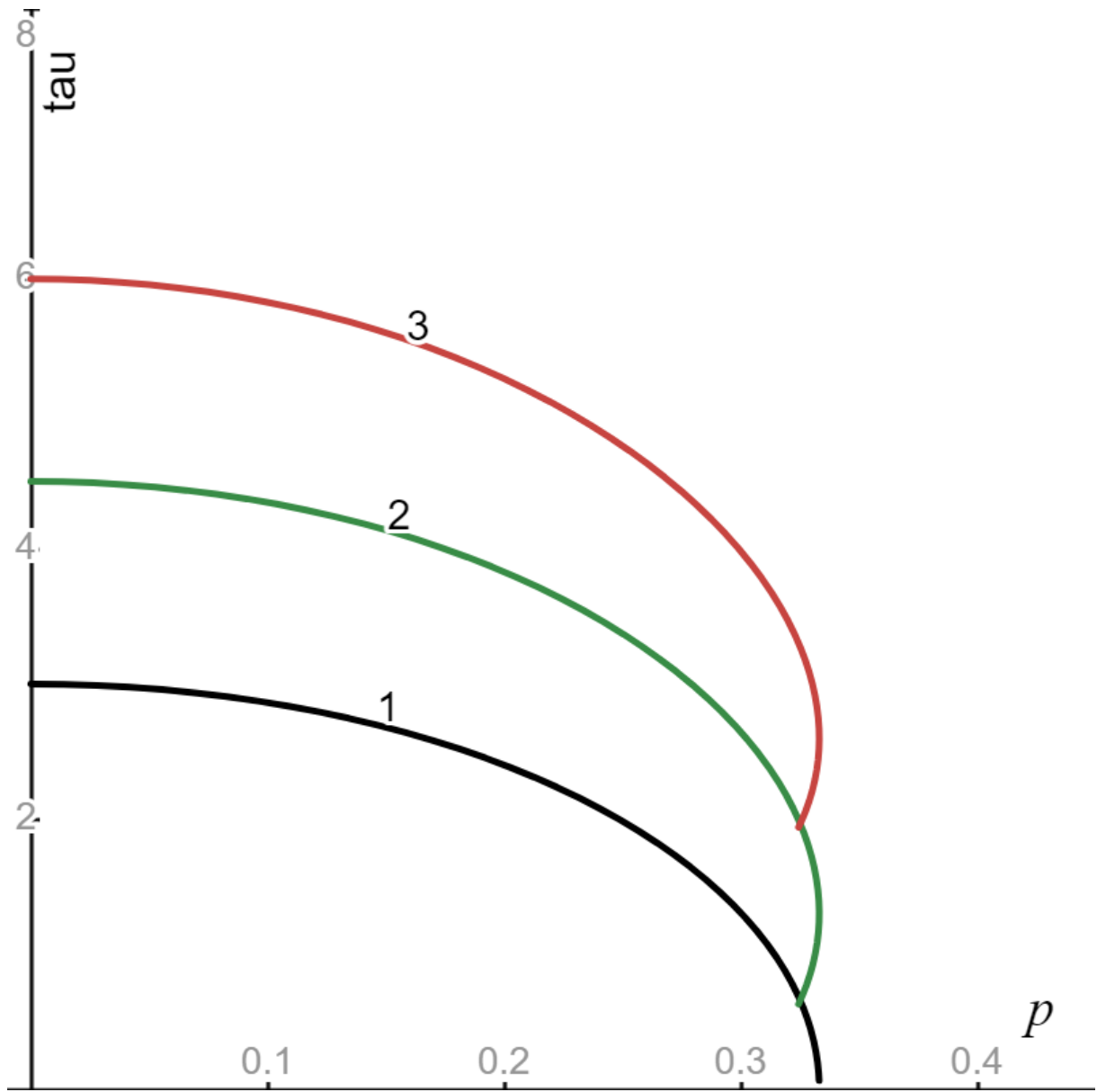

Figure 15: $\tau$-$p$ curves corresponding to primary (1) and its first (2) and second order (3) reverberations taking into account ray bending at medium-water column interface.

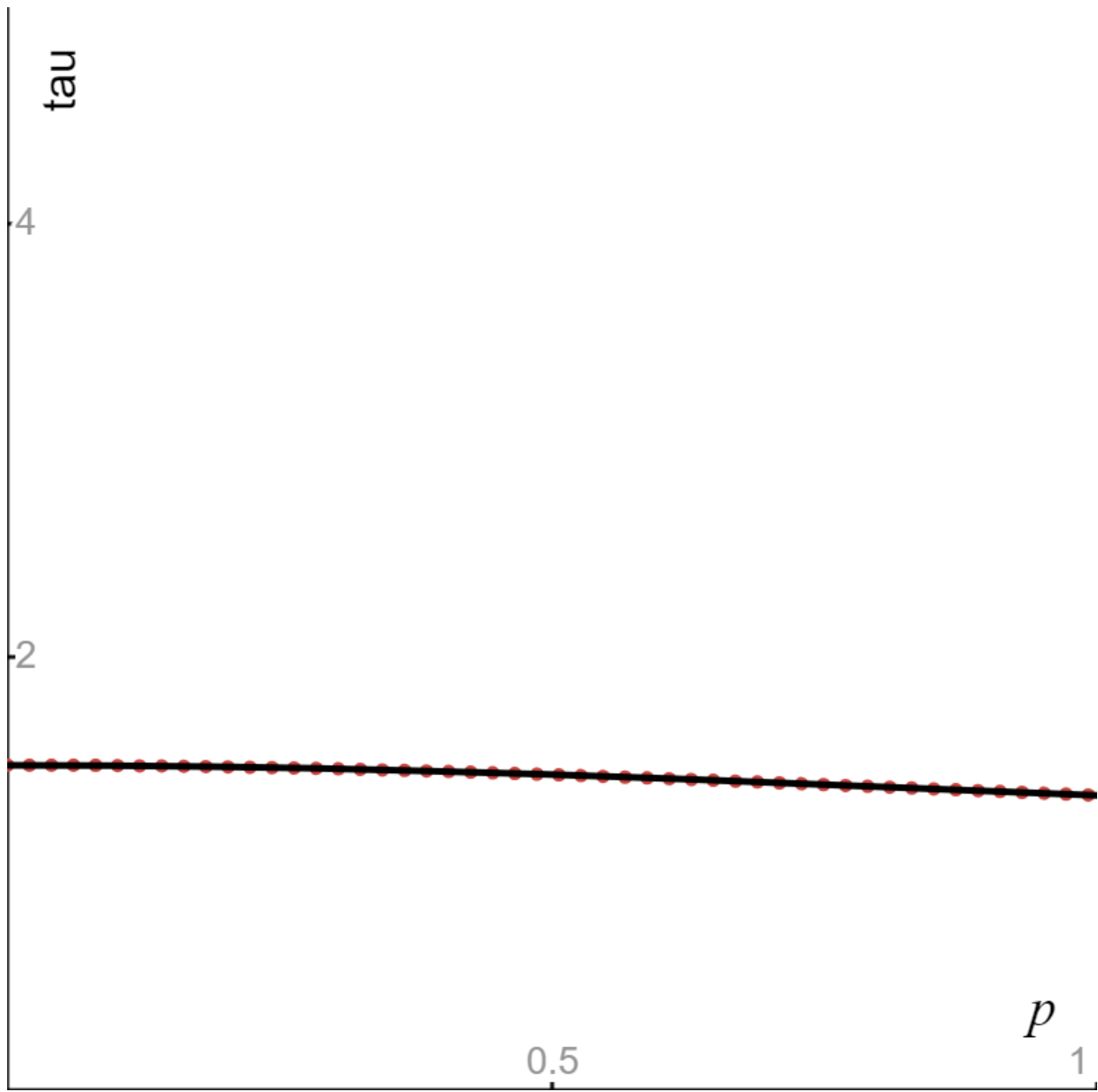

Figure 16: Differences between consecutive curves 1,2 and 3 of Figure 15 are plotted here. We see that curve (2)-curve (1) (dotted curve) is exactly equal to curve (3) – curve (2) (solid black curve), showing perfect periodicity.

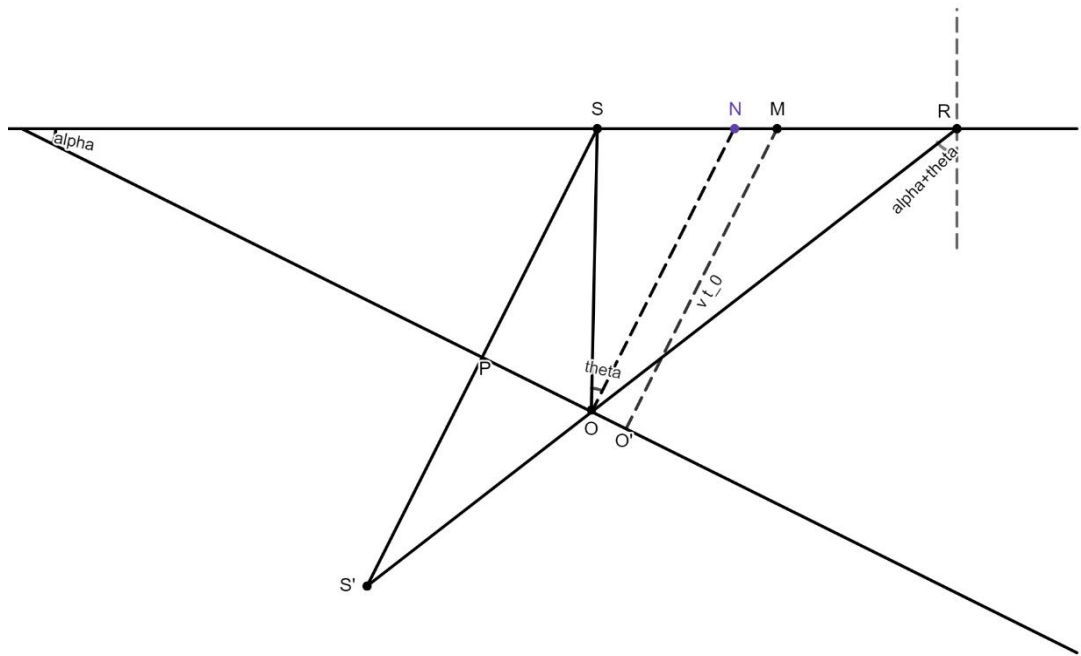

Figure 17: Case of a single dipping bed (with dip 'alpha') in the subsurface for source S, receiver R and angle of incidence 'theta' is considered here.